\documentclass[ALICE,manyauthors]{cernphprep}
\usepackage[comma,square,numbers,sort&compress]{natbib}
\usepackage{lineno}
\usepackage{xspace}
\usepackage{amsmath,amssymb}
\usepackage[comma,square,numbers,sort&compress]{natbib}

\usepackage{hyperref}

\usepackage{esvect}
\usepackage{color}
\usepackage[T1]{fontenc}
\begin{document}
%%%%%%%%%%%%%%%%%%%%%%%%%%%%%%%%%%%%%%%%%%%%%%%%%%
% These are some new commands that may be useful 
% for paper writing in general. If other newcommands
% are needed for your specific paper, please feel 
% free to add here. 
%
% The currently available commands are organized in: 
% 1) Systems
% 2) Quantities
% 3) Energies and units
% 4) Detectors
% 5) particle species 
%%%%%%%%%%%%%%%%%%%%%%%%%%%%%%%%%%%%%%%%%%%%%%%%%%

% 1) SYSTEMS 
\newcommand{\pp}           {pp\xspace}
\newcommand{\ppbar}        {\mbox{$\mathrm {p\overline{p}}$}\xspace}
\newcommand{\XeXe}         {\mbox{Xe--Xe}\xspace}
\newcommand{\PbPb}         {\mbox{Pb--Pb}\xspace}
\newcommand{\pA}           {\mbox{pA}\xspace}
\newcommand{\pPb}          {\mbox{p--Pb}\xspace}
\newcommand{\AuAu}         {\mbox{Au--Au}\xspace}
\newcommand{\dAu}          {\mbox{d--Au}\xspace}

% 2) QUANTITIES 
\newcommand{\s}            {\ensuremath{\sqrt{s}}\xspace}
\newcommand{\snn}          {\ensuremath{\sqrt{s_{\mathrm{NN}}}}\xspace}
\newcommand{\pt}           {\ensuremath{p_{\rm T}}\xspace}
\newcommand{\meanpt}       {$\langle p_{\mathrm{T}}\rangle$\xspace}
\newcommand{\ycms}         {\ensuremath{y_{\rm CMS}}\xspace}
\newcommand{\ylab}         {\ensuremath{y_{\rm lab}}\xspace}
\newcommand{\etarange}[1]  {\mbox{$\left | \eta \right |~<~#1$}}
\newcommand{\yrange}[1]    {\mbox{$\left | y \right |~<~#1$}}
\newcommand{\dndy}         {\ensuremath{\mathrm{d}N_\mathrm{ch}/\mathrm{d}y}\xspace}
\newcommand{\dndeta}       {\ensuremath{\mathrm{d}N_\mathrm{ch}/\mathrm{d}\eta}\xspace}
\newcommand{\avdndeta}     {\ensuremath{\langle\dndeta\rangle}\xspace}
\newcommand{\dNdy}         {\ensuremath{\mathrm{d}N_\mathrm{ch}/\mathrm{d}y}\xspace}
\newcommand{\Npart}        {\ensuremath{N_\mathrm{part}}\xspace}
\newcommand{\Ncoll}        {\ensuremath{N_\mathrm{coll}}\xspace}
\newcommand{\dEdx}         {\ensuremath{\textrm{d}E/\textrm{d}x}\xspace}
\newcommand{\RpPb}         {\ensuremath{R_{\rm pPb}}\xspace}

% 3) ENERGIES, UNITS
\newcommand{\nineH}        {$\sqrt{s}~=~0.9$~Te\kern-.1emV\xspace}
\newcommand{\seven}        {$\sqrt{s}~=~7$~Te\kern-.1emV\xspace}
\newcommand{\twoH}         {$\sqrt{s}~=~0.2$~Te\kern-.1emV\xspace}
\newcommand{\twosevensix}  {$\sqrt{s}~=~2.76$~Te\kern-.1emV\xspace}
\newcommand{\five}         {$\sqrt{s}~=~5.02$~Te\kern-.1emV\xspace}
\newcommand{\twosevensixnn}{$\sqrt{s_{\mathrm{NN}}}~=~2.76$~Te\kern-.1emV\xspace}
\newcommand{\fivenn}       {$\sqrt{s_{\mathrm{NN}}}~=~5.02$~Te\kern-.1emV\xspace}
\newcommand{\LT}           {L{\'e}vy-Tsallis\xspace}
\newcommand{\GeVc}         {Ge\kern-.1emV/$c$\xspace}
\newcommand{\MeVc}         {Me\kern-.1emV/$c$\xspace}
\newcommand{\TeV}          {Te\kern-.1emV\xspace}
\newcommand{\GeV}          {Ge\kern-.1emV\xspace}
\newcommand{\MeV}          {Me\kern-.1emV\xspace}
\newcommand{\GeVmass}      {Ge\kern-.2emV/$c^2$\xspace}
\newcommand{\MeVmass}      {Me\kern-.2emV/$c^2$\xspace}
\newcommand{\lumi}         {\ensuremath{\mathcal{L}}\xspace}

% 4) DETECTORS 
\newcommand{\ITS}          {\rm{ITS}\xspace}
\newcommand{\TOF}          {\rm{TOF}\xspace}
\newcommand{\ZDC}          {\rm{ZDC}\xspace}
\newcommand{\ZDCs}         {\rm{ZDCs}\xspace}
\newcommand{\ZNA}          {\rm{ZNA}\xspace}
\newcommand{\ZNC}          {\rm{ZNC}\xspace}
\newcommand{\SPD}          {\rm{SPD}\xspace}
\newcommand{\SDD}          {\rm{SDD}\xspace}
\newcommand{\SSD}          {\rm{SSD}\xspace}
\newcommand{\TPC}          {\rm{TPC}\xspace}
\newcommand{\TRD}          {\rm{TRD}\xspace}
\newcommand{\VZERO}        {\rm{V0}\xspace}
\newcommand{\VZEROA}       {\rm{V0A}\xspace}
\newcommand{\VZEROC}       {\rm{V0C}\xspace}
\newcommand{\Vdecay} 	   {\ensuremath{V^{0}}\xspace}

% 4) PARTICLE SPECIES 
\newcommand{\ee}           {\ensuremath{e^{+}e^{-}}} 
\newcommand{\pip}          {\ensuremath{\pi^{+}}\xspace}
\newcommand{\pim}          {\ensuremath{\pi^{-}}\xspace}
\newcommand{\kap}          {\ensuremath{\rm{K}^{+}}\xspace}
\newcommand{\kam}          {\ensuremath{\rm{K}^{-}}\xspace}
\newcommand{\pbar}         {\ensuremath{\rm\overline{p}}\xspace}
\newcommand{\kzero}        {\ensuremath{{\rm K}^{0}_{\rm{S}}}\xspace}
\newcommand{\lmb}          {\ensuremath{\Lambda}\xspace}
\newcommand{\almb}         {\ensuremath{\overline{\Lambda}}\xspace}
\newcommand{\Om}           {\ensuremath{\Omega^-}\xspace}
\newcommand{\Mo}           {\ensuremath{\overline{\Omega}^+}\xspace}
\newcommand{\X}            {\ensuremath{\Xi^-}\xspace}
\newcommand{\Ix}           {\ensuremath{\overline{\Xi}^+}\xspace}
\newcommand{\Xis}          {\ensuremath{\Xi^{\pm}}\xspace}
\newcommand{\Oms}          {\ensuremath{\Omega^{\pm}}\xspace}
\newcommand{\degree}       {\ensuremath{^{\rm o}}\xspace}
\newcommand{\Dmeson}       {\ensuremath{\rm D^{0}}\xspace}
\newcommand{\Dbar}       {\ensuremath{\rm\overline{D}{}^0}\xspace}
\newcommand{\DDbar}       {\ensuremath{\Dmeson~{\rm and} ~\Dbar}\xspace}

%%%%%%%%%%%%%%%  Title page %%%%%%%%%%%%%%%%%%%%%%%%
\begin{titlepage}
% the dates below correspond to CERN approval
% please don't touch: EB chairs will take care
\PHyear{2019}       % required, will be obtained from CERN
\PHnumber{241}      % required, will be obtained from CERN
\PHdate{28 October}  % required, will be obtained from CERN 
%%%%%%%%%%%%%%%%%%%%%%%%%%%%%%%%%%%%%%%%%%%%%%%%%%%%
%%TC:ignore
%%% Put your own title + short title here:
\title{Probing the effects of strong electromagnetic fields with charge-dependent directed flow in Pb--Pb collisions at the LHC}
\ShortTitle{Charge-dependent directed flow of hadrons and D mesons}   % appears on left page headers

%%% Do not change the next lines
\Collaboration{ALICE Collaboration\thanks{See Appendix~\ref{app:collab} for the list of collaboration members}}
\ShortAuthor{ALICE Collaboration} % appears on right page headers, do not change
%%TC:endignore

\begin{abstract}

The first measurement at the LHC of charge-dependent directed flow ($v_{1}$) relative to the spectator plane is presented for Pb--Pb collisions at $\sqrt{s_{\rm NN}}$ = 5.02~TeV.
Results are reported for charged hadrons and \Dmeson mesons for the transverse momentum intervals $p_{\rm T}>0.2$~GeV/$c$ and $3<p_{\rm T}<$ 6 GeV/$c$ in the 5--40\% and 10--40\% centrality classes, respectively.
The difference between the positively and negatively charged hadron $v_{1}$ has a positive slope as a function of pseudorapidity $\eta$, ${\rm d}\Delta{v_1}/{\rm d}\eta=$[1.68 $\pm$ 0.49 (stat.) $\pm$ 0.41 (syst.)] $\times 10^{-4}$. 
The same measurement for \Dmeson and \Dbar mesons yields a positive value ${\rm d}\Delta{v_1}/{\rm d}\eta$= [4.9 $\pm$ 1.7 (stat.) $\pm$ 0.6 (syst.)]$\times 10^{-1}$,
which is about three orders of magnitude larger than the one of the charged hadrons.
These measurements can provide new insights into the effects of the strong electromagnetic field and the initial tilt of matter created in non-central heavy-ion collisions on the dynamics of light (u, d, and s) and heavy (c) quarks.
The large difference between the observed $\Delta{v_1}$ of charged hadrons and \Dmeson mesons may reflect different sensitivity of the charm and light quarks to the early time dynamics of a heavy-ion collision.
These observations challenge some recent theoretical calculations, which predicted a negative and an order of magnitude smaller value of ${\rm d}\Delta{v_1}/{\rm d}\eta$ for both light-flavour and charmed hadrons.

\end{abstract}
\end{titlepage}

\setcounter{page}{2} %please do not remove this line

%%%%%%%%%%%%%%%%%%%%%%%%%%%%%%%%
% begin main text
%%%%%%%%%%%%%%%%%%%%%%%%%%%%%%%%

%\section{Introduction} 

Quantum Chromo-dynamic (QCD) calculations on the lattice~\cite{Bazavov:2018mes,Bazavov:2017dus,Bazavov:2014pvz, pQCD4,pQCD2,pQCD3} predict at high temperatures the existence of a deconfined state of quarks and gluons, known as the quark--gluon plasma (QGP).
Characterizing the QGP properties is among the main goals of the experimental program with ultra-relativistic heavy-ion collisions at the Large Hadron Collider (LHC).
Measurements of the anisotropic transverse flow~\cite{Aamodt:2010pa,ALICE:2011ab,ATLAS:2012at,Chatrchyan:2012ta,Abelev:2014pua}, quantified by the second ($v_2$) and higher order ($n>2$) harmonic coefficients $v_n$, allowed one to characterize the different phases of a heavy-ion collision and to constrain the properties of the QGP~\cite{Heinz:2013th,Gale:2013da,Niemi:2015qia,Bernhard:2016tnd,Dubla:2018czx}. 

The directed flow, $v_1$, has a special role due to its sensitivity to the three-dimensional spatial profile of the initial conditions and the pre-equilibrium early time dynamics in the evolution of the collision.
The space-momentum correlations in particle production from a longitudinally tilted source result in a non-zero $v_1$. The tilt arises from the asymmetries in the number of forward and backward moving participant nucleons at different positions in the transverse plane~\cite{Snellings:1999bt,Bozek:2010bi,Chatterjee:2017ahy}.
%Its origin is the space-momentum correlation in particle production given by the antisymmetric (rapidity odd) emission from the longitudinally tilted area of the initial entropy deposit formed by the forward and backward moving nucleons~\cite{Snellings:1999bt,Bozek:2010bi,Chatterjee:2017ahy}.
The directed flow of charged hadrons at the LHC~\cite{Abelev:2013cva} has significantly smaller magnitude compared to that at lower RHIC energies~\cite{Abelev:2008jga}, which can be interpreted as a smaller initial tilt at the LHC~\cite{Adamczyk:2016eux, Adil:2005bb, Adil:2005qn}.

Charm quarks are produced early in the collision via hard scattering processes.
Their emission region is not tilted in the longitudinal direction~\cite{Chatterjee:2017ahy} unlike the one of light quarks, which are predominantly produced in soft processes at later stages of the collision~\cite{Bozek:2010bi,Csernai:2011gg}.
Consequently, the region of charm quark production in the transverse plane is shifted with respect to that of light quarks and gluons, resulting in an enhanced dipole asymmetry in the charm quark distribution~\cite{Chatterjee:2017ahy}.
%This difference results in a shift of the initial transverse density profile of the charm quarks at non-zero rapidity relative to that of light quarks~\cite{Chatterjee:2017ahy}.
During the system expansion, charm quarks would be dragged by the flow of the light quarks in the transverse direction of the shift, which is predicted to result in a larger $v_1$ of charm hadrons compared to light-flavour hadrons~\cite{Chatterjee:2017ahy, Chatterjee:2018lsx}.
Consequently, the measurements of the charge-integrated directed flow of hadrons containing light (u, d, and s) and heavy (c) quarks together with their difference in magnitude are of great interest and allow one to probe the three-dimensional space-time evolution of the produced matter.

Heavy-ion collisions are also characterized by extremely strong electromagnetic fields primarily induced by spectator protons, which do not undergo inelastic collisions.
There is strong interest in characterizing the time evolution of these fields, which are estimated to reach $\rm 10^{18}$ -- $\rm 10^{19}$ Gauss in the early stages ($<$ 0.5 fm) of Pb--Pb collisions at LHC energies~\cite{Tuchin:2013ie,Skokov:2009qp}.
Phenomena predicted to occur in the presence of this strong electromagnetic field include the chiral magnetic effect (CME), which is driven by the generation of an electric current along the magnetic field in a medium with chiral imbalance~\cite{Fukushima:2008xe,Kharzeev:2007jp,Tuchin:2010vs,Voronyuk:2011jd}.
While experimental results for charge-dependent correlations are in qualitative
agreement with theoretical expectations for the CME~\cite{Khachatryan:2016got,Abelev:2012pa,Abelev:2009ac}, possible background contributions, such as effects of local charge conservation coupled with the anisotropic flow, prevent their unambiguous interpretation~\cite{Schlichting:2010qia} and have led to upper limits on the CME at LHC energies.
Thus it is fundamental to use other observables with direct sensitivity to the electromagnetic fields in order to constrain their magnitudes and time evolution in heavy-ion collisions.

The charge dependence of the produced particle directed flow relative to the spectator plane is directly sensitive to the presence of electromagnetic fields. The spectator plane is defined by the deflection direction of the collision spectators.
On average its orientation is perpendicular to the direction of the magnetic field generated by the positively charged spectators.
%In the center-of-mass system of two colliding nuclei 
The charge dependence of $v_1$ comes from two competing effects, acting in opposite directions, i.e. the Lorentz force and the Coulomb force, both induced by the rapidly decreasing magnetic field.
In an electrically conducting plasma the induced charged currents might slow down the decay of the magnetic field~\cite{Tuchin:2013ie}.
The measurement of charge-dependent $v_1$ can therefore constrain the QGP electric conductivity.

First estimates of these effects on the $v_1$ of charged particles were presented in~\cite{Rybicki:2013qla, Gursoy:2014aka}. 
Using the electric conductivity from lattice QCD calculations~\cite{Ding:2010ga, Amato:2013naa}, the difference between the $v_1$ of positively and negatively charged pions, $\Delta{}v_{1}(\pi) = v_{1}(\pi^+) - v_{1}(\pi^-)$, in Pb--Pb collisions at $\sqrt{s_{\rm NN}}$ = 2.76 TeV was estimated to be not larger than $10^{-5}$ for $|\eta|~<~1$~\cite{Gursoy:2014aka}.
Charm quarks, which are produced in the early stages when the magnetic field is maximal, should be more strongly affected by the electromagnetic fields than light quarks~\cite{Das:2016cwd, Chatterjee:2018lsx}.
The difference $\Delta{}v_{1}$ between the $v_1$ of \Dmeson($\rm{c\bar{u}}$) and \Dbar($\rm{\bar{c}u}$) mesons should therefore provide better sensitivity to this initial magnetic field. A calculation of $v_{1}$ of charmed mesons in the rapidity interval $|y|~<~1$ gives a value $\Delta{}v_{1}(D)~\sim~10^{-2}$~\cite{Das:2016cwd}, which is three orders of magnitude larger than the one expected for pions.
Recently, the STAR Collaboration published a measurement of directed flow of \Dmeson and \Dbar mesons in 10--80\% central Au--Au
collisions at $\sqrt{s_{\rm NN}}$ = 200 GeV~\cite{Adam:2019wnk}.
The slope of $v_1$  of \DDbar mesons as a function of rapidity is negative and about a factor of 25 times larger than
that of charged kaons. 

This Letter reports the first measurements at the LHC of the charge dependence of $v_1$ relative to the spectator plane for charged hadrons and \Dmeson mesons as a function of pseudorapidity in mid-central Pb--Pb collisions at $\sqrt{s_{\rm NN}}$ = 5.02 TeV.

About 23 (19) million Pb--Pb collisions in the 5--40\% (10--40\%) centrality interval are used for the charged hadron (\DDbar meson) $v_{1}$ measurements. Only events with a primary vertex reconstructed within $\pm$10 cm from the detector centre along the beam direction are analysed.
Two forward scintillator arrays (V0A and V0C)~\cite{vZero} are used to determine the collision centrality.
For the most central (0--5\%) collisions, the small number of spectators prevents an accurate reconstruction of their deflection.
In the 5--10\% centrality interval the large combinatorial background does not allow the measurement of the \DDbar $v_{1}$.

The deflection direction of the collision spectators is reconstructed from spectator neutrons detected using two Zero Degree Calorimeters (ZDC)~\cite{Aamodt:2008zz,bib::Alice_performance}.
The ZDCs have a 2$\times$2 segmentation in the plane transverse to the beam direction and are installed at 112.5~m distance from the detector centre on both sides of the interaction point, covering the "projectile" ($\eta>8.78$) and the "target" ($\eta<-8.78$) spectator regions.
For each ZDC a flow vector is constructed following the procedure described in~\cite{Abelev:2013cva}
\begin{equation}
{\rm\bf Q}^{ t,p} \equiv ({\rm Q}_x^{t,p}, {\rm Q}_y^{ t,p}) = \left.{\sum_{i = 1}^{4} {\rm\bf n}_{i} E^{ t,p}_i}\right/{\sum_{i = 1}^{4} E_{i}^{t,p}},
\label{Eq:Qvec}
\end{equation}
where $p$ and $t$ denote the ZDC on the projectile and target side, $E_i$ is the measured signal and ${\rm\bf n}_i$ = $(x_i,y_i)$ are the coordinates of the center of the $i$-th ZDC segment.

The deflection direction of the spectator neutrons is estimated event-by-event with the ${\rm\bf Q}^{ t,p}$ vectors corrected for the run-dependent variation of the LHC beam crossing position~\cite{Selyuzhenkov:2007zi}.
In mid-central collisions this deflection direction is strongly correlated with the magnetic field orientation.
The deflection is expected to be opposite (anti-correlated) for the projectile and the target sides, i.e. 
$\langle {\rm Q}_{x}^{p} {\rm Q}_{x}^{t} \rangle$ = $\langle {\rm Q}_{y}^{p} {\rm Q}_{y}^{t} \rangle$ $<$ 0 and $\langle {\rm Q}_{y}^{p} {\rm Q}_{x}^{t} \rangle$ and $\langle {\rm Q}_{x}^{p} {\rm Q}_{y}^{t} \rangle$ = 0. 
A deviation from these expectations, mostly for peripheral collisions with centrality above 40\%, is observed even after applying the flow-vector correction.
These residual variations are used in the estimation of the systematic uncertainty as described in~\cite{Abelev:2013cva} and discussed below.

The directed flow is measured using the scalar-product method~\cite{Luzum:2012da}
\begin{equation}
v_{1}^{t,p}
= \frac{\left\langle {\rm\bf u}{\rm\bf Q}^{ t,p} \right\rangle}{\sqrt{ \Big| \left\langle {\rm\bf Q}^{ t} {\rm\bf Q}^{ p} \right\rangle  \Big|}}
= \frac{\left\langle {\rm u_{x}} {\rm Q}_{x}^{ t,p} + {\rm u_{y}} {\rm Q}_{y}^{ t,p} \right\rangle}{\sqrt{ \Big| \left\langle {\rm Q}_{x}^{ t} {\rm Q}_{x}^{ p} + {\rm Q}_{y}^{ t} {\rm Q}_{y}^{ p} \right\rangle  \Big|}},
\label{Eq:v1AC}
\end{equation}
where ${\rm\bf u} = (\cos\varphi,\sin\varphi)$ is the unit flow vector of the charged hadron or \Dmeson meson candidate with azimuthal angle $\varphi$.
The directed flow is calculated as: $v_{1}$~=~($v_1^{ p}$~-~$v_1^{ t}$)/2.
The sign of $v_1$ is defined relative to the deflection of the projectile spectators, corresponding to the rapidity-odd component of the $v_1$ discussed in~\cite{Abelev:2013cva}.
The measurement of $v_1$ using spectators does not require any treatment of the momentum conservation unlike the measurements based on correlations between particles produced at midrapidity \cite{Retinskaya:2012ky}.
This is justified by the observation of a vanishing relative momentum shift along the spectator plane at $\eta = 0$ \cite{Abelev:2013cva}.

The charged-hadron $v_1$ is measured from tracks reconstructed with the Inner Tracking System (ITS)~\cite{Aamodt:2010aa} and the Time Projection Chamber (TPC) ~\cite{ALME2010316} and selected requiring $p_{\rm T}$ $>$ 0.2 GeV/$c$, $|\eta|~<~\rm 0.8$, at least 70 (out of a maximum of 159) TPC space points and $\chi^{2}$/ndf $<$ 2 for the momentum fit in the TPC. In order to reduce the contamination from secondary particles only tracks with a maximum distance of closest approach (DCA) to the reconstructed primary vertex in both the transverse ($\rm DCA_{xy}~<$ 2.4 cm) and the longitudinal direction ($\rm DCA_{z}~<$ 3.2 cm) are accepted.
%An average transverse momentum of the selected charged hadrons is $\langle p_{\rm T} \rangle \approx 0.7$~GeV/$c$.

The \Dmeson and \Dbar mesons are reconstructed using the decay channel $\Dmeson~\rightarrow~ K^{-} \pi^{+}$ and its charge conjugate for $3~<~p_{\rm T} <~6$~GeV/$c$.
%, which corresponds to a $\langle p_{\rm T} \rangle \approx 4.2$~GeV/$c$ calculated as described in~\cite{ALICE:2011aa}.
%The $\langle p_{\rm T} \rangle$ of the selected \Dmeson mesons is larger than that of the charged hadrons considered for the $v_1$ measurement. 
Pions and kaons are reconstructed in the TPC and ITS detectors.
Tracks are selected requiring $|\eta|~<$ 0.8, $p_{\rm T}~>$ 0.4 GeV/$c$, at least 70 hits in TPC and at least 2 hits (out of a maximum of 6) in the ITS, out of which at least one has to be in the two innermost layers.
Particle identification is based on measurements of the specific ionisation energy loss d$E$/d$x$ in the TPC and the flight time from the interaction point to the Time-Of-Flight (TOF) detector \cite{Carnesecchi:2018oss}.
%Tracks not matched to a hit in the TOF are identified using only the TPC information.
The charge of the identified pions and kaons allows one to distinguish between the $\Dmeson~\rightarrow~ K^{-} \pi^{+}$ and $\Dbar~\rightarrow~ K^{+} \pi^{-}$ candidates.
Geometrical selections on the displaced decay vertex topology are applied to reduce the combinatorial background~\cite{Acharya:2017qps}.
%The main selection variables are the separation between the primary and decay vertices, the displacement of the decay tracks from the primary vertex and the pointing of the reconstructed D-meson momentum to the primary vertex.

The $v_1^{\rm D}$ is extracted separately for \DDbar mesons via a simultaneous fit to the number $N(M)$ of $K^{\mp} \pi^{\pm}$ pairs and their $v_{1}(M)$ as a function of the invariant mass, $M$:
\begin{eqnarray}
N(M) &=& N_D(M)+N_{\rm bg}(M),
\label{Eq:N_Fit} \\
v_1(M) &=& [v^{D}_{1} N_D(M) + v^{\rm bg}_{1}(M) N_{\rm bg}(M)]/[N_D(M)+N_{\rm bg}(M)].
\label{Eq:v1_Fit}
\end{eqnarray}

\begin{figure}[ht!]
\centering
\includegraphics[width=0.45\textwidth]{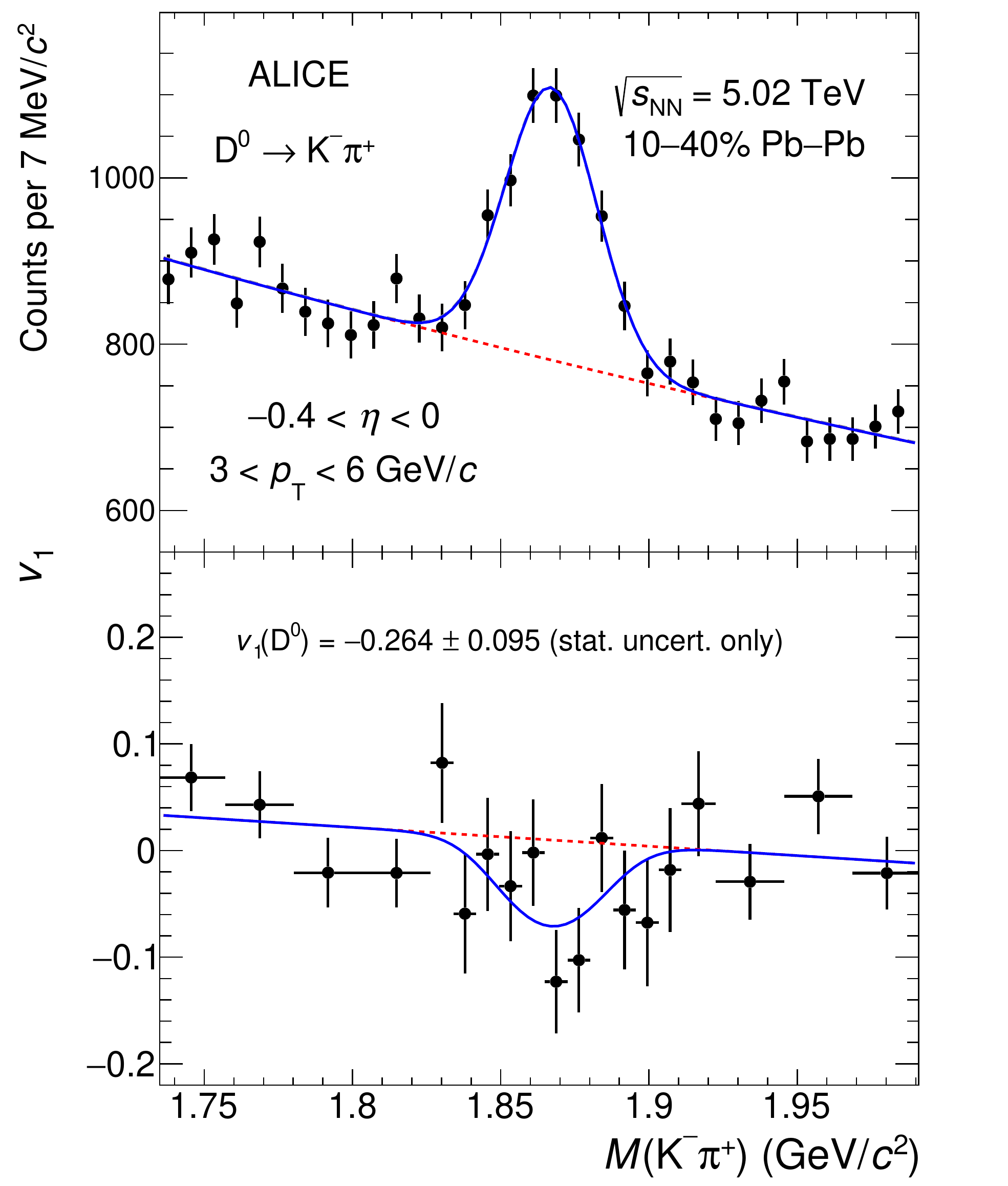}
\caption{
(color online)
Illustration of the \Dmeson meson $v_{1}$ extraction procedure via a simultaneous fit to the candidate invariant mass distribution (upper panel) and $v_{1}$ (lower panel) for $3 < p_{\rm T} < 6$~GeV/$c$ and $-0.4<\eta<0$ in the 10--40\% centrality interval.
The blue solid lines correspond to the combined signal and background fit functions, while the red dashed lines represent the background contribution.
%Coarse invariant mass intervals are used in the lower panel to reduce statistical fluctuations of the background $v_{1}$.
}
\label{Fig:fit}
\end{figure}

An example of the simultaneous fit is shown in Fig.~\ref{Fig:fit}.
The invariant mass distribution is fitted with the sum of a Gaussian function $N_D(M)$ for the \DDbar signal and an exponential function $N_{\rm bg}(M)$ for the background.
The invariant mass dependence of the directed flow of background candidates $v^{\rm bg}_{1}(M)$ is parametrized by a linear function.

Candidates, which satisfy both the $\rm K^{-} \pi^{+}$ and $\rm K^{+} \pi^{-}$ hypotheses (reflected kinematics) and therefore can not be tagged uniquely as \Dmeson or \Dbar, are rejected.
This removes about 35\% of the signal and increases the signal-to-background ratio by about 30--40\%, with a net result of a negligible reduction of the statistical significance of the \DDbar yield.
%Moreover, the inclusion of the reflected candidates requires to introduce in Eq.~\ref{Eq:v1_Fit} an additional fit parameter to account for their $v_1$ and to use the shape of their invariant mass distribution extracted from the GEANT Monte Carlo simulations of the ALICE detector response~\cite{Brun:1994aa}.
%Fits with these additional terms did not yield a stable fit result.
The extracted $v_1^{\rm D}$ includes contributions from both prompt \Dmeson mesons and feed-down \Dmeson mesons from beauty-hadron decays.
The fraction of prompt \Dmeson meson is about 85\% for the analysed centrality class and $p_{\rm T}$ interval~\cite{Acharya:2018hre}.

Common sources of systematic uncertainty between charged hadrons and D mesons are related to the resolution of the spectator plane and to the dependence on the ALICE magnet polarity.
The absolute systematic uncertainty related to the residual asymmetry in the spectator plane estimation is given by the difference between the $v_1$ obtained separately from $\langle{\rm u_{x}} {\rm Q}_{x}\rangle$ and $\langle{\rm u_{y}} {\rm Q}_{y}\rangle$ correlations with the ZDCs in Eq.~\ref{Eq:v1AC}.
It is about $3.5\times 10^{-5}$ ($2\times 10^{-2}$) for charged hadrons (\DDbar mesons).
Effects related to track reconstruction and geometrical alignment of the detectors, which could influence positive and negative tracks differently, are estimated by comparing the $v_1$ results obtained using data taken with opposite magnet polarity.
This comparison also probes the bias in the spectator plane estimation due to the non-zero beam crossing angle in the vertical plane, which had opposite values ($\pm$60 $\mu$rad) for the opposite magnet polarities.
The absolute difference between the $v_1$ values obtained with the two field polarities is ${\rm 2.5 \times 10^{-5}}$ ($2\times 10^{-2}$) for charged hadrons (\DDbar).
These systematic uncertainties are correlated in pseudorapidity for charged hadrons, while for \DDbar mesons no significant correlation, beyond statistical uncertainties, is observed.

For charged hadrons, the track quality selections are varied and an absolute systematic uncertainty of ${\rm 2.5 \times 10^{-5}}$ is assigned.
The contribution from secondaries is varied by changing the maximum $\rm DCA_{xy}$, which resulted in a negligible variation of $v_1$.
The contamination due to TPC tracks originating from pile-up collisions during the readout time of the TPC is estimated by varying the selections on the correlations between the event multiplicity (centrality) estimated with detectors with different readout times. The resulting systematic uncertainty is ${\rm 10^{-5}}$ for the charged hadrons.
No systematic uncertainty is assigned for \DDbar because the topological selections effectively remove tracks from pile-up.
The uncertainty due to the \DDbar signal extraction is estimated by varying (i) the fit functions in Eq.~(\ref{Eq:N_Fit}) and (\ref{Eq:v1_Fit}) for $N(M)$ and $v_1(M)$, (ii) fixing the Gaussian width and mean to the values extracted from Monte Carlo simulations, and (iii) varying the invariant mass fit range.
The absolute systematic uncertainty assigned to $v_{1}$ due to the \DDbar yield extraction is $2\times 10^{-2}$.
The possible bias due to the $p_{\rm T}$-dependent efficiency in the \DDbar $v_1$ analysis is tested by re-weighting both signal and background with the inverse value of the signal reconstruction efficiency as a function of $p_{\rm T}$.
%An average $p_{\rm T}$ for the re-weighted results reduces by less than 6\%. 
The assigned absolute systematic uncertainty is $10^{-2}$.

The total systematic uncertainty on $v_{1}$ is obtained by adding in quadrature the contributions described above.
In the calculation of $\Delta v_{1}(D)$ all individual systematic uncertainties are propagated as fully uncorrelated between \Dmeson and \Dbar.
For charged hadrons, the systematic uncertainties due to the asymmetry in the spectator plane estimation and the magnet polarity are correlated between positive and negative tracks and largely cancel in $\Delta v_{1}(h)$.

\begin{figure}[ht!]
\centering
\includegraphics[width=1.\linewidth]{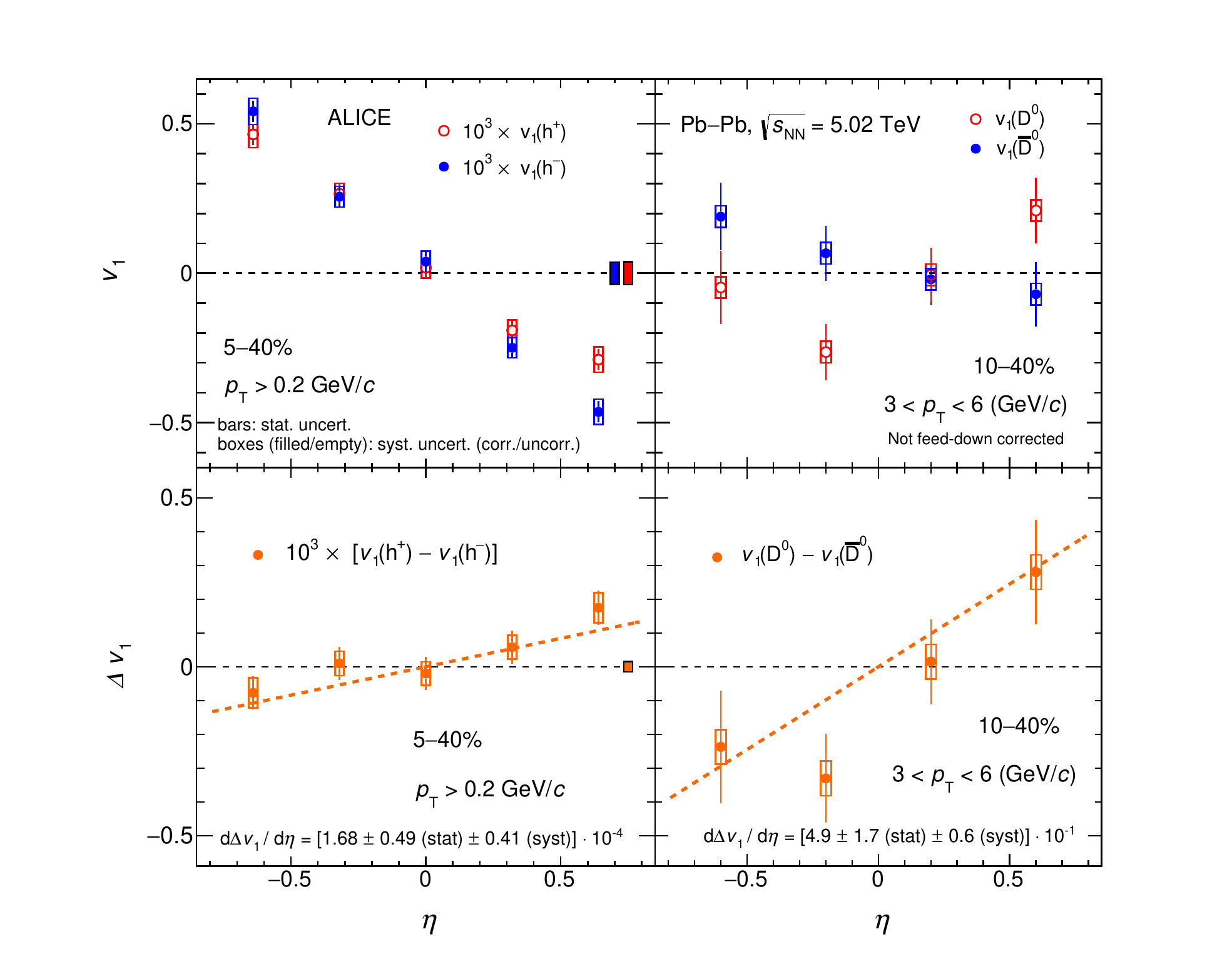}
\caption{
(color online)
Upper left: $v_{1}$ of positively (red) and negatively (blue) charged hadrons for the 5--40\% centrality interval.
Upper right: $v_{1}$ of \Dmeson (red) and \Dbar (blue) for the 10--40\% centrality interval. 
Lower left and right: $\Delta{}v_{1}(h) = v_{1}({h^+}) - v_{1}({h^-})$ and $\Delta{}v_{1}(\rm D) = v_{1}(\Dmeson) - v_{1}(\Dbar)$, respectively.
Dashed lines represent fits with a linear function.
}
 \label{v1all}
\end{figure}

The pseudorapidity dependence of the directed flow of positively and negatively charged hadrons for the 5--40\% centrality class in Pb--Pb collisions at $\sqrt{s_{\rm NN}}$ = 5.02 TeV is shown in the upper left panel of Fig.~\ref{v1all}.
The negative slope of $v_{1}$ is usually attributed to the effect of the initial tilt~\cite{Bozek:2010bi} or rotation~\cite{Csernai:2011gg} of the particle-emitting source.
The charge-integrated $v_{1}$ at $\sqrt{s_{\rm NN}}$ = 5.02 TeV agrees within uncertainties with the results at $\sqrt{s_{\rm NN}}$ = 2.76 TeV~\cite{Abelev:2013cva}.

The difference $\Delta{}v_{1}(h)$ between the $v_{1}$ of positively and negatively charged hadrons as a function of pseudorapidity is shown in the lower left panel of Fig.~\ref{v1all}.
The rapidity slope d$\Delta v_1/{\rm d}\eta$, extracted with a linear fit (constrained to $v_1$ = 0 at $\eta$ = 0), is d$\Delta v_1/{\rm d}\eta$ = [1.68 $\pm$ 0.49 (stat.) $\pm$ 0.41 (syst.)] $\times$ $\rm 10^{-4}$ with a significance of 2.6$\sigma$ for having a positive value.
The d$\Delta v_1/{\rm d}\eta$ is expected to reflect different effects, including those originating from the early-time magnetic field dynamics~\cite{Das:2016cwd, Chatterjee:2018lsx, Chatterjee:2017ahy}, the Coulomb interaction with charged spectators~\cite{Gursoy:2018yai}, as well as the transport to mid-rapidity via the baryon stopping mechanism~\cite{Snellings:1999bt} of the positive charge carried by the protons from the colliding nuclei.
The importance of baryon stopping for the charge dependence of unidentified hadron $v_1$ is supported by the observed difference, even at top RHIC energy, between proton and anti-proton $v_1$~\cite{Adamczyk:2011aa,Adamczyk:2014ipa,Adamczyk:2016eux}.
The baryon stopping effects are expected to decrease with increasing collision energy, as supported by the observation of a smaller magnitude of $v_1$~\cite{Abelev:2013cva} and of a proton-to-antiproton ratio closer to unity at the LHC as compared to RHIC~\cite{Abelev:2013vea}.
Despite the overall decrease, the baryon stopping can contribute significantly to the proton and anti-proton $v_1$ difference, and as such to the charge dependence of the inclusive hadron $v_1$.
%The role of baryon stopping can be clarified with future high precision measurements at the LHC.

The charged-hadron d$\Delta v_1/{\rm d}\eta$ at $\sqrt{s_{\rm NN}}$ = 5.02 TeV is one order of magnitude larger and has an opposite sign with respect to calculations for charged pions at $\sqrt{s_{\rm NN}}$ = 2.76 TeV~\cite{Gursoy:2014aka} based on the analytic solution of relativistic hydrodynamics~\cite{Gubser:2010ze} with a constant electrical conductivity of the QGP.
More recent calculations~\cite{Gursoy:2018yai}, using viscous hydrodynamic calculations~\cite{Shen:2014vra}, yield an absolute value of d$\Delta v_1/{\rm d}\eta$ of similar magnitude as the one measured for charged hadrons, but with opposite sign.

The \Dmeson and \Dbar $v_1$  as a function of pseudorapidity is shown in the upper right panel of Fig.~\ref{v1all}.
The data suggest a positive slope for the rapidity dependence of the $v_{1}$ of \Dmeson and a negative slope for \Dbar, with a significance of about 2$\sigma$ in both cases.
The slopes are different from the measurements in Au--Au collisions at $\sqrt{s_{\rm NN}}$ = 200 GeV~\cite{Adam:2019wnk}, where a negative value is observed for both \Dmeson and \Dbar.
Additionally, the $v_1$ for \DDbar mesons 
with 3 $<$ $p_{\rm T}$ $<$ 6 GeV/$c$ ($\langle p_{\rm T} \rangle$ $\approx$ 4.2 GeV/$c$) in the 10--40\% centrality interval is about three orders of magnitude larger than that of charged hadrons with $p_{\rm T}$ $>$ 0.2 GeV/$c$ ($\langle p_{\rm T} \rangle$ $\approx$ 0.7 GeV/$c$) in the 5--40\% centrality class. 
The different $p_{\rm T}$ intervals used for the charged-hadron and $\rm D^{0}$-meson $v_1$ measurements are imposed by the statistical precision of the data, which simultaneously limits the yield of high-$p_{\rm T} $ charged hadrons and results in low significance of the \DDbar meson yield at low $p_{\rm T}$.
The charged hadron $v_1$ at the LHC has a weak centrality dependence and changes sign around $p_{\rm T}\approx 1.5$~GeV/$c$~\cite{Abelev:2013cva}.
The differences in centrality and transverse momentum intervals should not be responsible for the observed difference between the magnitude of the $v_1$ of charged hadrons and \DDbar mesons.
The \DDbar $v_1$ is an order of magnitude larger than the predictions from the transport~\cite{Das:2016cwd} and hydrodynamic~\cite{Chatterjee:2018lsx, Chatterjee:2017ahy} model calculations.
The difference between the $v_{1}$ values of \Dmeson and \Dbar mesons $\Delta{}v_{1}(D)$ is shown in the lower right panel of Fig.~\ref{v1all}.
The value of d$\Delta v_1/\rm{d}\eta$ = [4.9 $\pm$ 1.7 (stat.) $\pm$ 0.6 (syst.)] $\times 10^{-1}$ corresponds to a significance of 2.7$\sigma$ to have a positive slope.
A negative value for d$\Delta v_1/\rm{d}\eta$ was predicted in~\cite{Das:2016cwd} and is observed in Au--Au collisions at $\sqrt{s_{\rm NN}}$ = 200 GeV~\cite{Adam:2019wnk}.
The opposite sign of the measured $\rm D^{0}$-meson and charged-hadron $\Delta v_{1}$ slope with respect to model calculations might indicate a stronger effect of the Lorentz force relative to the Coulomb one. These results demonstrate the sensitivity of the $v_{1}$ to the interplay among the effects of the rapidly decreasing magnetic field and the initial tilt of the source.

In summary, first measurements of the charge dependence of $v_{1}$ relative to the spectator plane in mid-central Pb--Pb collisions at $\sqrt{s_{\rm NN}}$ = 5.02 TeV are presented. 
The $v_{1}$ and the difference $\Delta v_1$ between positively and negatively charged hadrons and \Dmeson mesons are sensitive to the effects of the electromagnetic fields induced by spectator protons, baryon number transport, and the initial tilt or rotation of the particle-emitting source for non-central collisions.
An indication of a positive slope d$\Delta{}v_{1}/{\rm d}\eta$ of the charge-dependent $v_{1}$ at midrapidity for both charged hadrons and \DDbar mesons is observed.
The slope d$\Delta{}v_{1}/{\rm d}\eta$ is [1.68 $\pm$ 0.49 (stat.) $\pm$ 0.41 (syst.)] $\times 10^{-4}$ for charged hadrons with $p_{\rm T} > 0.2$~GeV/$c$ and [4.9 $\pm$ 1.7 (stat.) $\pm$ 0.6 (syst.)]$\times 10^{-1}$ for \DDbar mesons with $3<p_{\rm T}<6$~GeV/$c$, with significance of $2.6\sigma$ and $2.7\sigma$ for having a positive value, respectively.
The measured values of $v_1$ for \Dmeson and \Dbar mesons are about three orders of magnitude larger than the one of charged hadrons. 
These measurements together with those at RHIC~\cite{Adam:2019wnk} provide new insights and can constrain the theoretical modelling~\cite{Das:2016cwd, Gursoy:2014aka} of electromagnetic effects.
Further constraints will be set by future higher precision measurements at the LHC~\cite{Citron:2018lsq, ALICE-PUBLIC-2019-001}. % and by exploring new observables~\cite{Tuchin:2010gx}.

%%%%%%%%%%%%%%%%%%%%%%%%%%%%%%%%
% end main text 
%%%%%%%%%%%%%%%%%%%%%%%%%%%%%%%%

%%%%% acknowledgements - handled by EB chairs 
\newenvironment{acknowledgement}{\relax}{\relax}
\begin{acknowledgement}
\section*{Acknowledgements}
% add specific acknowledgements here 
% ...but please don't remove the line below: funding agencies
% will be acknowledged with a custom tex file handled by EB chairs after Collab Round 2
% Version: 2019-10-12

The ALICE Collaboration would like to thank all its engineers and technicians for their invaluable contributions to the construction of the experiment and the CERN accelerator teams for the outstanding performance of the LHC complex.
The ALICE Collaboration gratefully acknowledges the resources and support provided by all Grid centres and the Worldwide LHC Computing Grid (WLCG) collaboration.
The ALICE Collaboration acknowledges the following funding agencies for their support in building and running the ALICE detector:
A. I. Alikhanyan National Science Laboratory (Yerevan Physics Institute) Foundation (ANSL), State Committee of Science and World Federation of Scientists (WFS), Armenia;
Austrian Academy of Sciences, Austrian Science Fund (FWF): [M 2467-N36] and Nationalstiftung f\"{u}r Forschung, Technologie und Entwicklung, Austria;
Ministry of Communications and High Technologies, National Nuclear Research Center, Azerbaijan;
Conselho Nacional de Desenvolvimento Cient\'{\i}fico e Tecnol\'{o}gico (CNPq), Financiadora de Estudos e Projetos (Finep), Funda\c{c}\~{a}o de Amparo \`{a} Pesquisa do Estado de S\~{a}o Paulo (FAPESP) and Universidade Federal do Rio Grande do Sul (UFRGS), Brazil;
Ministry of Education of China (MOEC) , Ministry of Science \& Technology of China (MSTC) and National Natural Science Foundation of China (NSFC), China;
Ministry of Science and Education and Croatian Science Foundation, Croatia;
Centro de Aplicaciones Tecnol\'{o}gicas y Desarrollo Nuclear (CEADEN), Cubaenerg\'{\i}a, Cuba;
Ministry of Education, Youth and Sports of the Czech Republic, Czech Republic;
The Danish Council for Independent Research | Natural Sciences, the VILLUM FONDEN and Danish National Research Foundation (DNRF), Denmark;
Helsinki Institute of Physics (HIP), Finland;
Commissariat \`{a} l'Energie Atomique (CEA), Institut National de Physique Nucl\'{e}aire et de Physique des Particules (IN2P3) and Centre National de la Recherche Scientifique (CNRS) and R\'{e}gion des  Pays de la Loire, France;
Bundesministerium f\"{u}r Bildung und Forschung (BMBF) and GSI Helmholtzzentrum f\"{u}r Schwerionenforschung GmbH, Germany;
General Secretariat for Research and Technology, Ministry of Education, Research and Religions, Greece;
National Research, Development and Innovation Office, Hungary;
Department of Atomic Energy Government of India (DAE), Department of Science and Technology, Government of India (DST), University Grants Commission, Government of India (UGC) and Council of Scientific and Industrial Research (CSIR), India;
Indonesian Institute of Science, Indonesia;
Centro Fermi - Museo Storico della Fisica e Centro Studi e Ricerche Enrico Fermi and Istituto Nazionale di Fisica Nucleare (INFN), Italy;
Institute for Innovative Science and Technology , Nagasaki Institute of Applied Science (IIST), Japanese Ministry of Education, Culture, Sports, Science and Technology (MEXT) and Japan Society for the Promotion of Science (JSPS) KAKENHI, Japan;
Consejo Nacional de Ciencia (CONACYT) y Tecnolog\'{i}a, through Fondo de Cooperaci\'{o}n Internacional en Ciencia y Tecnolog\'{i}a (FONCICYT) and Direcci\'{o}n General de Asuntos del Personal Academico (DGAPA), Mexico;
Nederlandse Organisatie voor Wetenschappelijk Onderzoek (NWO), Netherlands;
The Research Council of Norway, Norway;
Commission on Science and Technology for Sustainable Development in the South (COMSATS), Pakistan;
Pontificia Universidad Cat\'{o}lica del Per\'{u}, Peru;
Ministry of Science and Higher Education and National Science Centre, Poland;
Korea Institute of Science and Technology Information and National Research Foundation of Korea (NRF), Republic of Korea;
Ministry of Education and Scientific Research, Institute of Atomic Physics and Ministry of Research and Innovation and Institute of Atomic Physics, Romania;
Joint Institute for Nuclear Research (JINR), Ministry of Education and Science of the Russian Federation, National Research Centre Kurchatov Institute, Russian Science Foundation and Russian Foundation for Basic Research, Russia;
Ministry of Education, Science, Research and Sport of the Slovak Republic, Slovakia;
National Research Foundation of South Africa, South Africa;
Swedish Research Council (VR) and Knut \& Alice Wallenberg Foundation (KAW), Sweden;
European Organization for Nuclear Research, Switzerland;
Suranaree University of Technology (SUT), National Science and Technology Development Agency (NSDTA) and Office of the Higher Education Commission under NRU project of Thailand, Thailand;
Turkish Atomic Energy Agency (TAEK), Turkey;
National Academy of  Sciences of Ukraine, Ukraine;
Science and Technology Facilities Council (STFC), United Kingdom;
National Science Foundation of the United States of America (NSF) and United States Department of Energy, Office of Nuclear Physics (DOE NP), United States of America.
\end{acknowledgement}

%%%%%%%% Bibliography 
\bibliographystyle{utphys}% or utphys
\bibliography{biblio}
%%%%%%%%%%%%%%%%%%%%%%%%%%%%%%%%
% Appendices: yours (if any) + authorlist
%%%%%%%%%%%%%%%%%%%%%%%%%%%%%%%%
\newpage
\appendix
  
%
%\input{} % put your appendices here (if any)
%

%%%%% Authorlist - please do not touch: handled by EB chairs 
\section{The ALICE Collaboration}
\label{app:collab}
% Collaboration: CERN-LHC-ALICE
% Generation Date is 2019-10-13

% How to use:
%%%%%%%%% appendix with author list
%\appendix
%\section{The ALICE Collaboration}
%\label{app:collab}
%\input{Alice_Authorslist_XXXX-Axx-XX.tex}
\begingroup
\small
\begin{flushleft}
S.~Acharya\Irefn{org141}\And 
D.~Adamov\'{a}\Irefn{org94}\And 
A.~Adler\Irefn{org74}\And 
J.~Adolfsson\Irefn{org80}\And 
M.M.~Aggarwal\Irefn{org99}\And 
G.~Aglieri Rinella\Irefn{org33}\And 
M.~Agnello\Irefn{org30}\And 
N.~Agrawal\Irefn{org10}\textsuperscript{,}\Irefn{org53}\And 
Z.~Ahammed\Irefn{org141}\And 
S.~Ahmad\Irefn{org16}
% ...
\And
\renewcommand\labelenumi{\textsuperscript{\theenumi}~}

\section*{Affiliation notes}
\renewcommand\theenumi{\roman{enumi}}
\begin{Authlist}
\item \Adef{org*}Deceased
\item \Adef{orgI}Dipartimento DET del Politecnico di Torino, Turin, Italy
\item \Adef{orgII}M.V. Lomonosov Moscow State University, D.V. Skobeltsyn Institute of Nuclear, Physics, Moscow, Russia
\item \Adef{orgIII}Department of Applied Physics, Aligarh Muslim University, Aligarh, India
\item \Adef{orgIV}Institute of Theoretical Physics, University of Wroclaw, Poland
\end{Authlist}

\section*{Collaboration Institutes}
\renewcommand\theenumi{\arabic{enumi}~}
\begin{Authlist}
\item \Idef{org1}A.I. Alikhanyan National Science Laboratory (Yerevan Physics Institute) Foundation, Yerevan, Armenia
\item \Idef{org2}Bogolyubov Institute for Theoretical Physics, National Academy of Sciences of Ukraine, Kiev, Ukraine
\item \Idef{org3}Bose Institute, Department of Physics  and Centre for Astroparticle Physics and Space Science (CAPSS), Kolkata, India
\item \Idef{org4}Budker Institute for Nuclear Physics, Novosibirsk, Russia
\item \Idef{org5}California Polytechnic State University, San Luis Obispo, California, United States
\item \Idef{org6}Central China Normal University, Wuhan, China
\item \Idef{org7}Centre de Calcul de l'IN2P3, Villeurbanne, Lyon, France
\item \Idef{org8}Centro de Aplicaciones Tecnol\'{o}gicas y Desarrollo Nuclear (CEADEN), Havana, Cuba
\item \Idef{org9}Centro de Investigaci\'{o}n y de Estudios Avanzados (CINVESTAV), Mexico City and M\'{e}rida, Mexico
\item \Idef{org10}Centro Fermi - Museo Storico della Fisica e Centro Studi e Ricerche ``Enrico Fermi', Rome, Italy
\item \Idef{org11}Chicago State University, Chicago, Illinois, United States
\item \Idef{org12}China Institute of Atomic Energy, Beijing, China
\item \Idef{org13}Comenius University Bratislava, Faculty of Mathematics, Physics and Informatics, Bratislava, Slovakia
\item \Idef{org14}COMSATS University Islamabad, Islamabad, Pakistan
\item \Idef{org15}Creighton University, Omaha, Nebraska, United States
\item \Idef{org16}Department of Physics, Aligarh Muslim University, Aligarh, India
\item \Idef{org17}Department of Physics, Pusan National University, Pusan, Republic of Korea
\item \Idef{org18}Department of Physics, Sejong University, Seoul, Republic of Korea
\item \Idef{org19}Department of Physics, University of California, Berkeley, California, United States
\item \Idef{org20}Department of Physics, University of Oslo, Oslo, Norway
\item \Idef{org21}Department of Physics and Technology, University of Bergen, Bergen, Norway
\item \Idef{org22}Dipartimento di Fisica dell'Universit\`{a} 'La Sapienza' and Sezione INFN, Rome, Italy
\item \Idef{org23}Dipartimento di Fisica dell'Universit\`{a} and Sezione INFN, Cagliari, Italy
\item \Idef{org24}Dipartimento di Fisica dell'Universit\`{a} and Sezione INFN, Trieste, Italy
\item \Idef{org25}Dipartimento di Fisica dell'Universit\`{a} and Sezione INFN, Turin, Italy
\item \Idef{org26}Dipartimento di Fisica e Astronomia dell'Universit\`{a} and Sezione INFN, Bologna, Italy
\item \Idef{org27}Dipartimento di Fisica e Astronomia dell'Universit\`{a} and Sezione INFN, Catania, Italy
\item \Idef{org28}Dipartimento di Fisica e Astronomia dell'Universit\`{a} and Sezione INFN, Padova, Italy
\item \Idef{org29}Dipartimento di Fisica `E.R.~Caianiello' dell'Universit\`{a} and Gruppo Collegato INFN, Salerno, Italy
\item \Idef{org30}Dipartimento DISAT del Politecnico and Sezione INFN, Turin, Italy
\item \Idef{org31}Dipartimento di Scienze e Innovazione Tecnologica dell'Universit\`{a} del Piemonte Orientale and INFN Sezione di Torino, Alessandria, Italy
\item \Idef{org32}Dipartimento Interateneo di Fisica `M.~Merlin' and Sezione INFN, Bari, Italy
\item \Idef{org33}European Organization for Nuclear Research (CERN), Geneva, Switzerland
\item \Idef{org34}Faculty of Electrical Engineering, Mechanical Engineering and Naval Architecture, University of Split, Split, Croatia
\item \Idef{org35}Faculty of Engineering and Science, Western Norway University of Applied Sciences, Bergen, Norway
\item \Idef{org36}Faculty of Nuclear Sciences and Physical Engineering, Czech Technical University in Prague, Prague, Czech Republic
\item \Idef{org37}Faculty of Science, P.J.~\v{S}af\'{a}rik University, Ko\v{s}ice, Slovakia
\item \Idef{org38}Frankfurt Institute for Advanced Studies, Johann Wolfgang Goethe-Universit\"{a}t Frankfurt, Frankfurt, Germany
\item \Idef{org39}Fudan University, Shanghai, China
\item \Idef{org40}Gangneung-Wonju National University, Gangneung, Republic of Korea
\item \Idef{org41}Gauhati University, Department of Physics, Guwahati, India
\item \Idef{org42}Helmholtz-Institut f\"{u}r Strahlen- und Kernphysik, Rheinische Friedrich-Wilhelms-Universit\"{a}t Bonn, Bonn, Germany
\item \Idef{org43}Helsinki Institute of Physics (HIP), Helsinki, Finland
\item \Idef{org44}High Energy Physics Group,  Universidad Aut\'{o}noma de Puebla, Puebla, Mexico
\item \Idef{org45}Hiroshima University, Hiroshima, Japan
\item \Idef{org46}Hochschule Worms, Zentrum  f\"{u}r Technologietransfer und Telekommunikation (ZTT), Worms, Germany
\item \Idef{org47}Horia Hulubei National Institute of Physics and Nuclear Engineering, Bucharest, Romania
\item \Idef{org48}Indian Institute of Technology Bombay (IIT), Mumbai, India
\item \Idef{org49}Indian Institute of Technology Indore, Indore, India
\item \Idef{org50}Indonesian Institute of Sciences, Jakarta, Indonesia
\item \Idef{org51}INFN, Laboratori Nazionali di Frascati, Frascati, Italy
\item \Idef{org52}INFN, Sezione di Bari, Bari, Italy
\item \Idef{org53}INFN, Sezione di Bologna, Bologna, Italy
\item \Idef{org54}INFN, Sezione di Cagliari, Cagliari, Italy
\item \Idef{org55}INFN, Sezione di Catania, Catania, Italy
\item \Idef{org56}INFN, Sezione di Padova, Padova, Italy
\item \Idef{org57}INFN, Sezione di Roma, Rome, Italy
\item \Idef{org58}INFN, Sezione di Torino, Turin, Italy
\item \Idef{org59}INFN, Sezione di Trieste, Trieste, Italy
\item \Idef{org60}Inha University, Incheon, Republic of Korea
\item \Idef{org61}Institut de Physique Nucl\'{e}aire d'Orsay (IPNO), Institut National de Physique Nucl\'{e}aire et de Physique des Particules (IN2P3/CNRS), Universit\'{e} de Paris-Sud, Universit\'{e} Paris-Saclay, Orsay, France
\item \Idef{org62}Institute for Nuclear Research, Academy of Sciences, Moscow, Russia
\item \Idef{org63}Institute for Subatomic Physics, Utrecht University/Nikhef, Utrecht, Netherlands
\item \Idef{org64}Institute of Experimental Physics, Slovak Academy of Sciences, Ko\v{s}ice, Slovakia
\item \Idef{org65}Institute of Physics, Homi Bhabha National Institute, Bhubaneswar, India
\item \Idef{org66}Institute of Physics of the Czech Academy of Sciences, Prague, Czech Republic
\item \Idef{org67}Institute of Space Science (ISS), Bucharest, Romania
\item \Idef{org68}Institut f\"{u}r Kernphysik, Johann Wolfgang Goethe-Universit\"{a}t Frankfurt, Frankfurt, Germany
\item \Idef{org69}Instituto de Ciencias Nucleares, Universidad Nacional Aut\'{o}noma de M\'{e}xico, Mexico City, Mexico
\item \Idef{org70}Instituto de F\'{i}sica, Universidade Federal do Rio Grande do Sul (UFRGS), Porto Alegre, Brazil
\item \Idef{org71}Instituto de F\'{\i}sica, Universidad Nacional Aut\'{o}noma de M\'{e}xico, Mexico City, Mexico
\item \Idef{org72}iThemba LABS, National Research Foundation, Somerset West, South Africa
\item \Idef{org73}Jeonbuk National University, Jeonju, Republic of Korea
\item \Idef{org74}Johann-Wolfgang-Goethe Universit\"{a}t Frankfurt Institut f\"{u}r Informatik, Fachbereich Informatik und Mathematik, Frankfurt, Germany
\item \Idef{org75}Joint Institute for Nuclear Research (JINR), Dubna, Russia
\item \Idef{org76}Korea Institute of Science and Technology Information, Daejeon, Republic of Korea
\item \Idef{org77}KTO Karatay University, Konya, Turkey
\item \Idef{org78}Laboratoire de Physique Subatomique et de Cosmologie, Universit\'{e} Grenoble-Alpes, CNRS-IN2P3, Grenoble, France
\item \Idef{org79}Lawrence Berkeley National Laboratory, Berkeley, California, United States
\item \Idef{org80}Lund University Department of Physics, Division of Particle Physics, Lund, Sweden
\item \Idef{org81}Nagasaki Institute of Applied Science, Nagasaki, Japan
\item \Idef{org82}Nara Women{'}s University (NWU), Nara, Japan
\item \Idef{org83}National and Kapodistrian University of Athens, School of Science, Department of Physics , Athens, Greece
\item \Idef{org84}National Centre for Nuclear Research, Warsaw, Poland
\item \Idef{org85}National Institute of Science Education and Research, Homi Bhabha National Institute, Jatni, India
\item \Idef{org86}National Nuclear Research Center, Baku, Azerbaijan
\item \Idef{org87}National Research Centre Kurchatov Institute, Moscow, Russia
\item \Idef{org88}Niels Bohr Institute, University of Copenhagen, Copenhagen, Denmark
\item \Idef{org89}Nikhef, National institute for subatomic physics, Amsterdam, Netherlands
\item \Idef{org90}NRC Kurchatov Institute IHEP, Protvino, Russia
\item \Idef{org91}NRC «Kurchatov Institute»  - ITEP, Moscow, Russia
\item \Idef{org92}NRNU Moscow Engineering Physics Institute, Moscow, Russia
\item \Idef{org93}Nuclear Physics Group, STFC Daresbury Laboratory, Daresbury, United Kingdom
\item \Idef{org94}Nuclear Physics Institute of the Czech Academy of Sciences, \v{R}e\v{z} u Prahy, Czech Republic
\item \Idef{org95}Oak Ridge National Laboratory, Oak Ridge, Tennessee, United States
\item \Idef{org96}Ohio State University, Columbus, Ohio, United States
\item \Idef{org97}Petersburg Nuclear Physics Institute, Gatchina, Russia
\item \Idef{org98}Physics department, Faculty of science, University of Zagreb, Zagreb, Croatia
\item \Idef{org99}Physics Department, Panjab University, Chandigarh, India
\item \Idef{org100}Physics Department, University of Jammu, Jammu, India
\item \Idef{org101}Physics Department, University of Rajasthan, Jaipur, India
\item \Idef{org102}Physikalisches Institut, Eberhard-Karls-Universit\"{a}t T\"{u}bingen, T\"{u}bingen, Germany
\item \Idef{org103}Physikalisches Institut, Ruprecht-Karls-Universit\"{a}t Heidelberg, Heidelberg, Germany
\item \Idef{org104}Physik Department, Technische Universit\"{a}t M\"{u}nchen, Munich, Germany
\item \Idef{org105}Politecnico di Bari, Bari, Italy
\item \Idef{org106}Research Division and ExtreMe Matter Institute EMMI, GSI Helmholtzzentrum f\"ur Schwerionenforschung GmbH, Darmstadt, Germany
\item \Idef{org107}Rudjer Bo\v{s}kovi\'{c} Institute, Zagreb, Croatia
\item \Idef{org108}Russian Federal Nuclear Center (VNIIEF), Sarov, Russia
\item \Idef{org109}Saha Institute of Nuclear Physics, Homi Bhabha National Institute, Kolkata, India
\item \Idef{org110}School of Physics and Astronomy, University of Birmingham, Birmingham, United Kingdom
\item \Idef{org111}Secci\'{o}n F\'{\i}sica, Departamento de Ciencias, Pontificia Universidad Cat\'{o}lica del Per\'{u}, Lima, Peru
\item \Idef{org112}St. Petersburg State University, St. Petersburg, Russia
\item \Idef{org113}Stefan Meyer Institut f\"{u}r Subatomare Physik (SMI), Vienna, Austria
\item \Idef{org114}SUBATECH, IMT Atlantique, Universit\'{e} de Nantes, CNRS-IN2P3, Nantes, France
\item \Idef{org115}Suranaree University of Technology, Nakhon Ratchasima, Thailand
\item \Idef{org116}Technical University of Ko\v{s}ice, Ko\v{s}ice, Slovakia
\item \Idef{org117}Technische Universit\"{a}t M\"{u}nchen, Excellence Cluster 'Universe', Munich, Germany
\item \Idef{org118}The Henryk Niewodniczanski Institute of Nuclear Physics, Polish Academy of Sciences, Cracow, Poland
\item \Idef{org119}The University of Texas at Austin, Austin, Texas, United States
\item \Idef{org120}Universidad Aut\'{o}noma de Sinaloa, Culiac\'{a}n, Mexico
\item \Idef{org121}Universidade de S\~{a}o Paulo (USP), S\~{a}o Paulo, Brazil
\item \Idef{org122}Universidade Estadual de Campinas (UNICAMP), Campinas, Brazil
\item \Idef{org123}Universidade Federal do ABC, Santo Andre, Brazil
\item \Idef{org124}University of Cape Town, Cape Town, South Africa
\item \Idef{org125}University of Houston, Houston, Texas, United States
\item \Idef{org126}University of Jyv\"{a}skyl\"{a}, Jyv\"{a}skyl\"{a}, Finland
\item \Idef{org127}University of Liverpool, Liverpool, United Kingdom
\item \Idef{org128}University of Science and Technology of China, Hefei, China
\item \Idef{org129}University of South-Eastern Norway, Tonsberg, Norway
\item \Idef{org130}University of Tennessee, Knoxville, Tennessee, United States
\item \Idef{org131}University of the Witwatersrand, Johannesburg, South Africa
\item \Idef{org132}University of Tokyo, Tokyo, Japan
\item \Idef{org133}University of Tsukuba, Tsukuba, Japan
\item \Idef{org134}Universit\'{e} Clermont Auvergne, CNRS/IN2P3, LPC, Clermont-Ferrand, France
\item \Idef{org135}Universit\'{e} de Lyon, Universit\'{e} Lyon 1, CNRS/IN2P3, IPN-Lyon, Villeurbanne, Lyon, France
\item \Idef{org136}Universit\'{e} de Strasbourg, CNRS, IPHC UMR 7178, F-67000 Strasbourg, France, Strasbourg, France
\item \Idef{org137}Universit\'{e} Paris-Saclay Centre d'Etudes de Saclay (CEA), IRFU, D\'{e}partment de Physique Nucl\'{e}aire (DPhN), Saclay, France
\item \Idef{org138}Universit\`{a} degli Studi di Foggia, Foggia, Italy
\item \Idef{org139}Universit\`{a} degli Studi di Pavia, Pavia, Italy
\item \Idef{org140}Universit\`{a} di Brescia, Brescia, Italy
\item \Idef{org141}Variable Energy Cyclotron Centre, Homi Bhabha National Institute, Kolkata, India
\item \Idef{org142}Warsaw University of Technology, Warsaw, Poland
\item \Idef{org143}Wayne State University, Detroit, Michigan, United States
\item \Idef{org144}Westf\"{a}lische Wilhelms-Universit\"{a}t M\"{u}nster, Institut f\"{u}r Kernphysik, M\"{u}nster, Germany
\item \Idef{org145}Wigner Research Centre for Physics, Budapest, Hungary
\item \Idef{org146}Yale University, New Haven, Connecticut, United States
\item \Idef{org147}Yonsei University, Seoul, Republic of Korea
\end{Authlist}
\endgroup
  
\end{document}